\documentclass[sigconf]{acmart}
\renewcommand\footnotetextcopyrightpermission[1]{}

\usepackage{subcaption}
\usepackage{caption} \usepackage{textcomp}
\usepackage[algo2e,ruled,vlined,linesnumbered]{algorithm2e}
\usepackage{pifont}
\usepackage{booktabs}
\usepackage{upquote}
\usepackage{listings}
\usepackage{comment}
\usepackage{todonotes}
\usepackage{fancyvrb}
\usepackage{graphicx}
\usepackage{multicol}
\usepackage{multirow}
\usepackage{cleveref}
\usepackage{lstautogobble}  
\usepackage{color}          
\usepackage{zi4}            
\usepackage[breakable,skins]{tcolorbox}
\usepackage{listings}
\usepackage{xcolor}
\usepackage{diagbox}
\usepackage{booktabs}
\usepackage{colortbl}
\usepackage{bbding}
\usepackage{graphicx}
\usepackage{rotating}
\usepackage{adjustbox}
\usepackage{lipsum}
\usepackage{enumitem}
\usepackage{framed}
\usepackage[misc]{ifsym}

\definecolor{codegreen}{rgb}{0,0.6,0}
\definecolor{codegray}{rgb}{0.5,0.5,0.5}
\definecolor{codepurple}{rgb}{0.58,0,0.82}
\definecolor{backcolour}{rgb}{1,1,1}
\definecolor{assistantcolor}{RGB}{19,118,188}
\definecolor{usercolor}{RGB}{229,91,43}
\definecolor{criticcolor}{RGB}{150,115,166}

\newcommand{\mkchange}[1]{{#1}}

\tcbset{
  fullbox/.style={
    width=1\textwidth,
    colback=black!05,
    colframe=black!20,
    colbacktitle=black!50,
    enhanced,
    center,
    segmentation style={solid, draw=gray}
  }
}
\tcbset{
  shortbox/.style={
    width=0.45\textwidth,
    colback=black!05,
    colframe=black!20,
    colbacktitle=black!50,
    enhanced,
    center,
    segmentation style={solid, draw=gray}
  }
}
\newtcolorbox{ShortBox}[1][]{shortbox,before upper={\parindent1pt},#1}
\newtcolorbox{FullBox}[1][]{fullbox,before upper={\parindent1pt},#1}
\renewcommand{\thefootnote}{\fnsymbol{footnote}}

\begin{document}
\begin{sloppypar}
\title{Experimenting a New Programming Practice with LLMs}

\author{Simiao Zhang$^*$}
\email{smzhang@stu.ecnu.edu.cn}
\affiliation{%
  \institution{East China Normal University}
  \city{Shanghai}
  \country{China}
}

\author{Jiaping Wang$^*$}
\email{51265902031@stu.ecnu.edu.cn}
\affiliation{%
  \institution{East China Normal University}
  \city{Shanghai}
  \country{China}
}

\author{Guoliang Dong \textsuperscript{\Letter}}
\email{gldong@smu.edu.sg}
\affiliation{%
  \institution{Singapore Management University}
  \city{Singapore}
  \country{Singapore}
}

\author{Jun Sun}
\email{junsun@smu.edu.sg}
\affiliation{%
  \institution{Singapore Management University}
  \city{Singapore}
  \country{Singapore}
}

\author{Yueling Zhang}
\email{ylzhang@sei.ecnu.edu.cn}
\affiliation{%
  \institution{East China Normal University}
  \city{Shanghai}
  \country{China}
}
\author{Geguang Pu}
\email{ggpu@sei.ecnu.edu.cn}
\affiliation{%
  \institution{East China Normal University}
  \city{Shanghai}
  \country{China}
}

\begin{abstract}
The recent development on large language models makes automatically constructing small programs possible. It thus has the potential to free software engineers from low-level coding and allow us to focus on the perhaps more interesting parts of software development, such as requirement engineering and system testing. In this project, we develop a prototype named AISD (AI-aided Software Development), which is capable of taking high-level (potentially vague) user requirements as inputs, generates detailed use cases, prototype system designs, and subsequently system implementation. Different from existing attempts, AISD is designed to keep the user in the loop, i.e., by repeatedly taking user feedback on use  cases, high-level system designs, and prototype implementations through system testing. AISD has been evaluated with a novel benchmark of non-trivial software projects. The experimental results suggest that it might be possible to imagine a future where software engineering is reduced to requirement engineering and system testing only.
\end{abstract}

\keywords{Requirement engineering, system testing, large language model, code generation}

\maketitle
\def\thefootnote{*}\footnotetext{These authors contributed equally to this work}\def\thefootnote{\arabic{footnote}}
\def\thefootnote{\Letter}\footnotetext{Corresponding author}\def\thefootnote{\arabic{footnote}}

\section{Introduction}
Large language models (LLMs), i.e., transformer-based language models with a huge number of parameters, have shown remarkable performance in natural language understanding as well as solving complex problems thanks to their emergent abilities~\cite{emergent}. In particular, their abilities of instruction following, step-by-step reasoning, and in-context learning have led to many applications in a variety of domains, including code generation~\cite{zhao2023survey}. That is, given a description of a low-level simple coding task, LLMs such as GPT are capable of synthesizing programs automatically, often correctly too~\cite{humanEval,codellama,ni2023l2ceval}. Thus, it gives us a glint of hope that one day LLMs might free us from manually low-level coding. 

In fact, a few recent projects attempted the ambitious goal of replacing programmers with LLMs. 
Li et al.~\cite{chatDev} propose an end-to-end software development framework known as ChatDev, which mimics the classical waterfall model and breaks down the software development process into four stages, i.e., designing, coding, testing, and documenting. That is, given a high-level rather vague requirement, ChatDev leverages multiple LLM-based virtual roles to generate detailed requirements, design and implementation in that sequence through rounds of conversions. MetaGPT~\cite{metagpt} adopts a similar idea and further standardizes each LLM-based agent's output to guide the other agents in the subsequent tasks. For instance, during the design phase, MetaGPT generates Product Requirements Documents (PRDs) with a standardized structure to coordinate the subsequent development process. While these attempts are shown to improve the underlying LLM's performance to certain extent, they often fail when the software project is non-trivial~\cite{suri2023software}. 

This is hardly surprising as, even for experienced human programmers, it is infeasible to complete a complex software based on vague high-level requirements. 
This is precisely why requirement engineering and system testing play vital roles in the software development process, and software development processes such as rapid prototyping and agile methods value user feedback during system development highly. Requirement engineering and system testing are essential for eliciting the expectations of end-users and stakeholders, ensuring that the delivered software aligns with their expectations. Requirement engineering is a complex and multifaceted process, and no human beings can produce flawless requirement specifications in a single attempt. However, the above-mentioned approaches conduct requirement analysis only lightly and the identified requirements are directly passed on to the coding phase, depriving users of the opportunity to validate and modify the automatically generated requirements and implementation.

At the same time, it is perhaps fair to say that requirement engineering in the traditional programming paradigm failed to achieve its promises to some extent as well as it may often be disconnected from system implementation as the project development progresses, e.g., the requirement and corresponding system design documents often are not properly maintained along with the system implementation. One fundamental reason is that there is limited ways of obtaining timely feedback from system designers and programmers (since a software project often lasts months or even years). With the help of LLMs, we can potentially shorten the development time significantly, allow users to test/validate prototype implementations ``instantaneously'', collect user feedback (e.g., in the form of failed test cases or updated requirements) timely and refine the implementation accordingly. In other words, we can make requirement engineering more relevant and effective by timely testing the implementation. 

In this work, we experiment with a novel AI-powered software development framework called AISD. AISD distinguishes itself from existing approaches in two aspects. Firstly, AISD is designed to engage users throughout the software development process, especially during the requirement analysis, high-level system design and system validation phase. Secondly, AISD adheres to the philosophy that \emph{less is more} when engaging the human developers. Specifically, when presented with a vague requirement, AISD generates a requirement document (e.g., use cases) capturing only the core functions required by the system and one system design document describing which source files should be built, and seeks user-feedback. With the user-feedback, these documents are updated accordingly. Note that due to the limited attention span of the LLMs~\cite{tian2023chatgpt} (and humans too), both documents are designed to be simple but friendly for humans and LLMs, which we will show in Section~\ref{sec:3}. Subsequently, AISD decomposes the system design into low-level coding tasks and completes them systematically and automatically. Once a prototype is implemented (i.e., the resultant system passes the unit testing and basic system testing), users are engaged to validate the system to check whether their requirements are met. If any failures are identified, the implementation, the design and/or the requirements are updated accordingly to construct another prototype. This iterative process continues until the users accept the product. 

AISD has been implemented as a self-contained toolkit. Considering that existing benchmarks, e.g., HumanEval~\cite{humanEval}, MBPP~\cite{mbpp} and CAMEL~\cite{li2023camel}  are not suitable for evaluating the capability of the LLM-based software development frameworks (all of them either are limited to simple function-level implementation tasks or lack detailed requirements specifications for system-level implementation tasks), we have developed a novel benchmark named CAASD (\textbf{C}apability \textbf{A}ssessment of \textbf{A}utomatic \textbf{S}oftware \textbf{D}evelopment). Each task of CAASD is equipped with a list of reference use cases depicting the system requirements. The reference use cases are used to evaluate the quality and completeness of a system implementation. We have compared AISD with two state-of-the-art baselines ChatDev~\cite{chatDev} and MetaGPT~\cite{metagpt} on CAASD. The experimental results demonstrate that AISD achieves the highest pass rate while using the fewest tokens. On average, AISD achieves an impressive pass rate of \mkchange{75.2}\%. Relative to these two baselines, there are improvements of \mkchange{133.5}\% and \mkchange{501.6}\% respectively. 
Moreover, it reduces the number of tokens consumed from at least \mkchange{28734} to \mkchange{21993}. 
These experimental results provide compelling evidence for the importance of user engagement in AI-aided software development.

In summary, we make the following contributions.
\begin{itemize}
    \item We introduce AISD, an AI-aided software development framework that is designed to keep users engaged through requirement engineering and system testing. 
    \item  We build a novel benchmark named CAASD for objectively assessing the capabilities of various AI-aided software development systems. To the best of our knowledge, this is the first benchmark that offers criteria for assessing how well a software development task is completed.
    \item We conduct a comprehensive evaluation of AISD using the benchmark. The results underscore the critical role of human engagement during AI-powered software development. These results suggest a potential future where software engineering may be streamlined to focus primarily on requirement engineering.
    \vspace{-0.1cm}
\end{itemize}

The remainders of the paper are organized as follows. Section~\ref{sec:2} reviews some essential backgrounds. Section~\ref{sec:3} presents the detailed design of AISD. Section~\ref{sec:4} presents details on our experiments of applying AISD as well as two baselines to the CAASD benchmark. Section~\ref{sec:5} reviews related work and Section~\ref{sec:6} concludes. 

\begin{figure*}[t]
    \begin{subfigure}[t]{0.46\textwidth}
        \centering
        \begin{FullBox}
        A "whatpu" is a small, furry animal native to Tanzania. An example of a sentence that uses
        the word whatpu is:\\
        We were traveling in Africa and we saw these very cute whatpus.\\
        To do a "farduddle" means to jump up and down really fast. An example of a sentence that uses
        the word farduddle is:
        \tcblower
        \emph{(Output) When we won the game, we all started to farduddle in celebration.}
        \end{FullBox}
        \caption{Few-shot prompting}
        \label{fig:fewshot}
    \end{subfigure}
    \begin{subfigure}[t]{0.45\textwidth}
        \centering
        \begin{FullBox}
            Q: Roger has 5 balls. He buys 2 more cans of balls. Each can has 3 balls. How many balls does he have now?\\
            A: Roger started with 5 balls. 2 cans of 3 tennis balls each is 6 tennis balls. 5 + 6 = 11. The answer is 11\\
            Q: I had 23 apples. If I used 20 to make lunch and bought 6 more, how many apples do I have?
            \tcblower
            \emph{(Output) You started with 23 apples. You used 20 to make lunch. So you had 23 - 20 = 3. They bought 6 more apples, so you have 3 + 6 = 9. The answer is 9.}
        \end{FullBox}
        \caption{Few-shot CoT prompting}
        \label{fig:zeroshort}
    \end{subfigure}
    \caption{Examples of different prompting techniques}
\end{figure*}
\begin{figure*}
    \centering
    \includegraphics[width=0.7\textwidth]{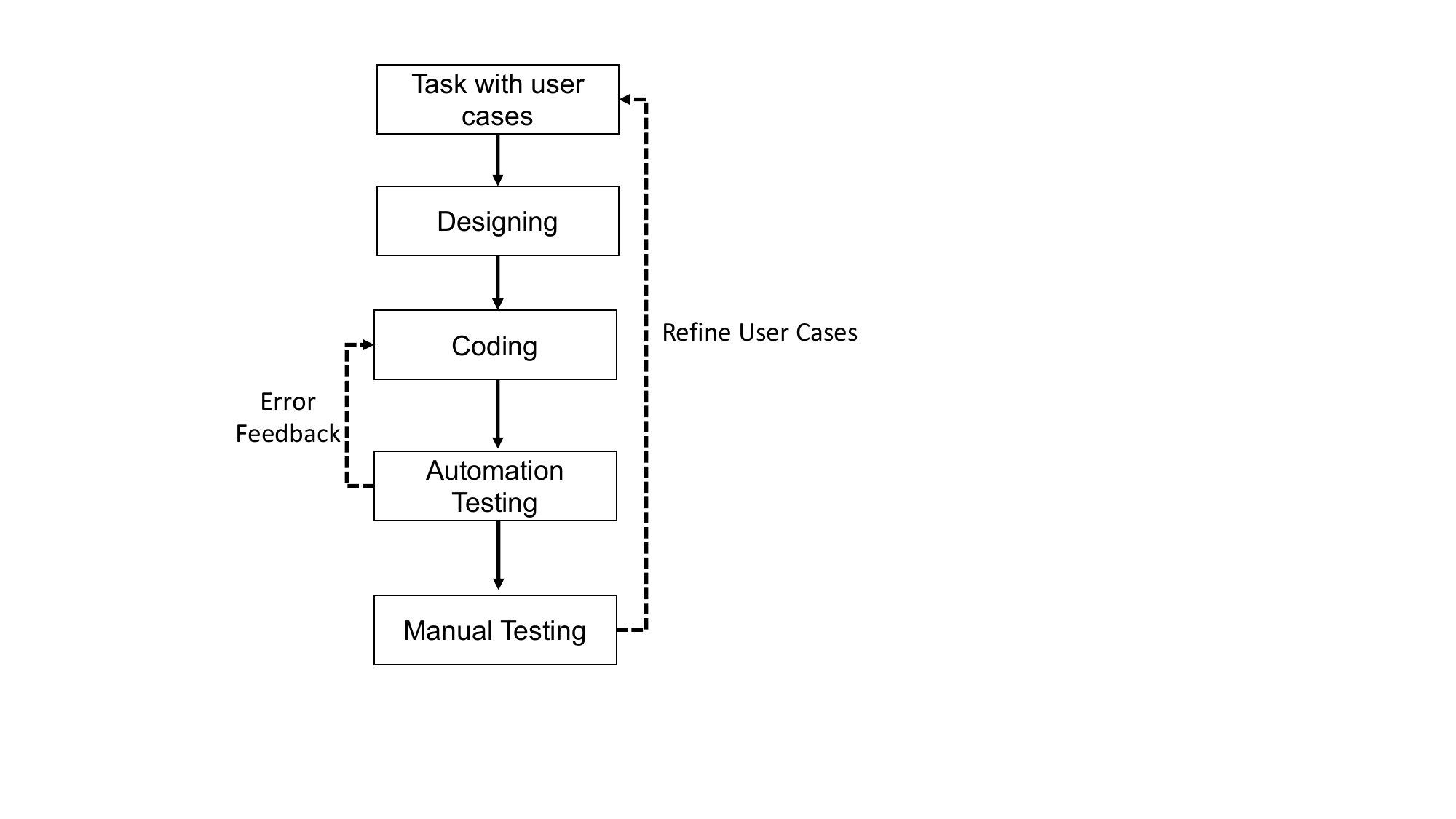}
    \caption{Overview of AISD}
    \label{fig:overview}
\end{figure*}
\vspace{-0.1cm}
\section{Preliminaries}
\label{sec:2}
In this section, we review relevant backgrounds on LLMs and prompt engineering.
\subsection{Large Language Models} LLMs to pre-trained language models~\cite{qiu2020pre} that have a heightened number of parameters, often in the range of tens or even thousands of billions~\cite{zhao2023survey}. LLMs, exemplified by models like GPT-3~\cite{floridi2020gpt} and PaLM~\cite{Palm}, outperform smaller-scale counterparts like BERT~\cite{bert} and GPT-2~\cite{gpt2} by not only achieving substantial performance improvements but also demonstrating emergent abilities. These emergent abilities, including in-context learning~\cite{icl}, instruction following~\cite{zeng2023evaluating}, and step-by-step reasoning~\cite{kojima2022large}, endow LLMs with the capability to tackle complex tasks. 

A prominent application showcasing the prowess of LLMs is ChatGPT, a significant chatbot that can complete various tasks from emulating human-like responses to aiding in debugging and writing programs. Such capabilities allow LLMs to potentially revolutionize many domains, with potential applications in software development and beyond. In this work, we utilize LLMs to experiment a new programming practice.

\subsection{Prompt Engineering}\label{prompt_engineering}
Prompt engineering is the practice of effectively exploiting and harnessing abilities of LLMs by optimizing prompts in a manner that LLMs can comprehend and interpret~\cite{qiao2022reasoning,liu2023pre}. A prompt is typically natural language text instructing how LLMs perform a task and specifying the desired output format. Prompt engineering is often based on the emergent abilities. That is, users can adopt certain prompt templates to have LLMs tackle complex and interesting tasks. In the following, we briefly introduce two prompting engineering techniques used in this work.

\emph{\textbf{Few-shot prompting.}} Few-shot prompting serves as a method to facilitate in-context learning by incorporating demonstrations within the prompt~\cite{brown2020language}. These demonstrations guide the model towards improved performance, acting as a form of conditioning for generating responses in subsequent examples. Figure~\ref{fig:fewshot} showcases an example of few-shot prompting. This example is from the work of Brown et al.~\cite{brown2020language}, the goal of which is to create a sentence using the given word. We can observe that the model completes this task successfully based on a single example. In this work, we use few-shot prompting to steer LLMs to outputs with certain format. 

\emph{\textbf{Chain-of-thought.}} Chain-of-thought (CoT) prompting~\cite{wei2022chain} is mainly based on the ``step-by-step reasoning'' ability (one of the emergent abilities) of LLMs. It is often combined with few-shot prompting to perform intricate tasks that require reasoning. In the prompt, users often need to demonstrate how they approach a similar task step by step. Figure~\ref{fig:zeroshort} is an example adapted from work~[\citenum{kojima2022large}]. In this example, the prompt exemplifies the calculation process, and the output indeed correctly follows it. Recently, Kojima et al. proposed zero-shot CoT prompting. Instead of exemplifying how to approach a task through examples, zero-shot CoT prompting simply adds ``Let’s think step by step'' after the question. In this work, we adopt the zero-shot CoT prompting in our prompts to improve outputs of LLMs.

\subsection{LLM-based Autonomous Agent}\label{llm_based_agent}
An LLM-based autonomous agent is a system that employs an LLM as its core controller to automatically plan tasks, make decisions, adopt actions, and reflect and update results. Technically, an LLM-based autonomous agent consists of four modules~\cite{wang2023survey}: profiling module, memory module, planning module and action module. 

The profiling module specifies the role of the agent by writing the related profile in the prompt, potentially inducing LLMs to produce response with the role's expertise. Different roles indicate different responsibilities for agents. For example, in the software development scenario, we may designate an agent as a programmer, focusing primarily on coding. The memory module is used to store past observations, decisions and actions, facilitating future actions. The planning module simulates the process that humans follow to handle a complex task. That is, this module focuses on breaking down a complex task to a series of simple and manageable subtasks. The action module is responsible for executing the agent's decisions (i.e., the decomposed subtasks). Note that the action module directly interacts with the environment, i.e., using external tools. LLM agents provide a flexible way to accomplish intricate tasks. Note that LLM agents have varying levels of autonomy, and some LLM agents may only have some of the four modules. 
Depending on the capabilities granted during the design phase, agents can exhibit self-directed behaviors ranging from purely reactive to highly proactive. In our work, AISD is mainly built with multiple communicative agents~\cite{li2023camel}, which complete their work through conversation with each other and human developers.
\vspace{-0.2cm}
\section{Our Approach}
\label{sec:3}
\begin{figure*}[t]
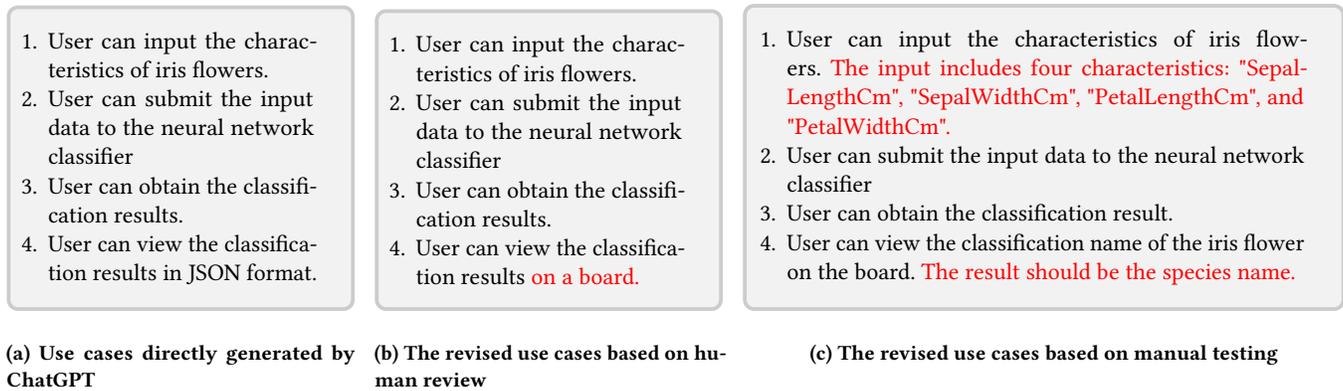

    \begin{subfigure}[t]{0.26\textwidth}
        \centering
        \begin{FullBox}
            \begin{enumerate}[leftmargin=0pt,label=\arabic*.]
              \item User can input the characteristics of iris flowers.
              \item User can submit the input data to the neural network classifier
              \item User can obtain the classification results.
              \item User can view the classification results in JSON format. 
            \end{enumerate}
          \end{FullBox}
        \caption{Use cases directly generated by ChatGPT}
        \label{fig:usecse1}
    \end{subfigure}
    \hfill
    \begin{subfigure}[t]{0.26\textwidth}
        \centering
        \begin{FullBox}
            \begin{enumerate}[leftmargin=0pt,label=\arabic*.]
              \item User can input the characteristics of iris flowers.
              \item User can submit the input data to the neural network classifier
              \item User can obtain the classification results.
              \item User can view the classification results \textcolor{red}{on a board.}
            \end{enumerate}
        \end{FullBox}
        \caption{The revised use cases based on human review}
        \label{fig:usecse2}
    \end{subfigure}
    \hfill
    \begin{subfigure}[t]{0.45\textwidth}
        \centering
        \begin{FullBox}
            \begin{enumerate}[leftmargin=1pt,label=\arabic*.]
                \item User can input the characteristics of iris flowers. \textcolor{red}{The input includes four characteristics: "SepalLengthCm", "SepalWidthCm", "PetalLengthCm", and "PetalWidthCm".}
                \item User can submit the input data to the neural network classifier
                \item User can obtain the classification result.
                \item User can view the classification name of the iris flower on the board. \textcolor{red}{The result should be the species name.}
            \end{enumerate}
        \end{FullBox}
        \caption{The revised use cases based on manual testing}
        \label{fig:usecse3}
    \end{subfigure}
    \caption{Use cases evolution of the task ``A tool for classifying iris flowers'' }
\end{figure*}

In this section, we present the design of AISD and its interaction with users. Figure~\ref{fig:overview} provides an overview of the overall workflow, where the tasks highlighted in green are the ones that rely on human-interaction. Starting with an initial idea, AISD firstly refines the task and generates a set of use cases to capture the core functions required by the desired system. Instead of immediately proceeding to the follow-up procedure, AISD interacts with the user to review and modify the generated use cases to ensure they properly convey the user's requirements. Once the user agrees to proceed, AISD produces a simple but LLM-and-human friendly system design. Note that humans can also revise the high-level system design as they do for use cases, but in practice, the revision of system design requires some expertise, which may be not suitable for ordinary users. We thus highlight this phase with light green indicating that human engagement in this phase is optional. After that, AISD implements the system automatically according to the design. Following implementation, AISD iteratively tests and refines the system. There are two types of testing. One is the automation testing and the other is the manual testing. When errors occur during automation testing, the error messages are used as feedback to guide the bug fixing automatically. Dashed lines in Figure~\ref{fig:overview} indicate that an iterative process is involved, and the red dashed lines indicate that humans are involved in the respective iteration. That is, AISD iteratively performs automation testing and bug fixing. To prevent AISD from getting stuck in this iteration (e.g., a bug that cannot be fixed by LLMs), AISD allows the user to set a maximum number of iteration times. After automation testing, unlike existing approaches that often terminate after code generation~\cite{li2023camel} or automation testing~\cite{chatDev,metagpt}, in AISD, the user is asked to test the resultant system manually. If a test failure occurs, different actions are taken according to the nature of the failure. Specifically, if there are any error messages reported during testing, these error messages are provided as input to the bug fixing module. Otherwise, the user may add or refine the use cases and instruct AISD to repeat the subsequent procedures (e.g., in the case of that some system requirement was missing or is revised (as often is the case in practice)). In the following, we introduce the details of each step. 

\subsection{Use Cases Generation}

Use cases play a crucial role in the software development. They are helpful in identifying, clarifying, and understanding the functional requirements of a system from a user's perspective. When implementing a desired system with LLMs, use cases can be integrated into specific prompts to articulate the concrete functions that should be implemented. Once a system is delivered, use cases can also serve as a means to validate whether the system meets the stack-holders' requirements. 

In the traditional software development process, use cases are typically derived from a general idea by requirements engineers. While some works, such as MetaGPT, show that LLMs can generate well-documented use cases in certain scenarios, we argue that involving stakeholders in the derivation of use cases is crucial due to the vagueness and incompleteness of requirements in the initial stage, particularly when tackling complex tasks. That is, automatically generated use cases may fail to clearly and correctly convey the functional requirements of the system. However, on the other hand, directly writing out the detailed use cases manually from scratch is usually challenging and burdensome for users. Therefore, to alleviate the burden the user bears and avoid fully relying on outputs of LLMs, AISD initially generates a ``draft'' of use cases using LLMs with the prompt shown in Figure~\ref{fig:pmusecase}, and then asks the user to review (and revises if necessary) the generated use cases. Note that each prompt used in AISD consists of two parts: the system message and the user message. The system message is used for role assignment and thus induces LLMs to produce response with the role’s expertise. The user message is used for task specification. Once the use cases are deemed acceptable, the user then instructs AISD to proceed with these use cases. 

For example, a user intends to develop a tool for identifying various species of iris. The user inputs the task ``develop a neural network classifier tool that allows users to input the characteristics of iris flowers and obtain classification results'' to AISD. AISD then generates an initial version of use cases as shown in Figure~\ref{fig:usecse1}. Although most of the use cases are acceptable, the last use case fails to capture the correct requirement as the user would like to view the result on a graphic interface, instead of in a JSON file. Consequently, the user revises the use case to align it with the desired requirement, as shown in Figure~\ref{fig:usecse2} where the revision is highlighted with red color. This use case review and refinement process may repeat multiple times if necessary until the user is happy with the resultant use cases. After that, the user instructs AISD to proceed.
\begin{figure}[t]
    \begin{ShortBox}
        \textbf{System Message:}\\
        You are a Product Manager. You have extensive experience in designing products and translating complex technical requirements into clear, user-centric scenarios.\\
        \textbf{User Message:}\\ According to the user's task listed below: Task: "\{task\}". You should write down the use cases required by this task. Output in JSON format. The format is:\\
        \{
            "1": "User can view the GUI."
        \}
    \end{ShortBox}
    \vspace{-0.3cm}
    \caption{Prompts used in use cases generation phase}
    \label{fig:pmusecase}
    \vspace{-0.6cm}
\end{figure}
\subsection{System Designing}
\label{sec:32}
\begin{figure}
    \begin{ShortBox}
    \textbf{System Message:}\\
    You are a software architect. According to the user's task and use cases, you will write the system design. List only key code files (no more than 6 files), ALWAYS start with the "main" file. Don't list multi-level files. Output in JSON format. The format is: \\
    \{"main.py": "This is the main file of ...", \}\\
    \textbf{User Message:}\\
    Task: \{task\}\\
    Use Cases: \{use\_cases\}
    \end{ShortBox}
    \caption{Prompts used for system designing}
    \label{fig:promDesign}
\end{figure}

        

        



        




System design serves as a blueprint for implementing a system. The standard system design process is complex, involving the design of architecture, components, modules, interfaces, and data to meet specified requirements~\cite{systemdesign}. Typically, this process generates numerous documents at different levels of abstraction. While these documents are beneficial for developers to understand and implement the desired system, our experience suggests that they are unsuitable for instructing LLMs to code accordingly due to two challenges. Firstly, LLMs struggle when processing comprehensive system design documents that exceed their token limit. Secondly, the intricate and overwhelming information in these documents often distracts LLMs from focusing on the core task—implementing functional requirements. To address these challenges, we propose a simplified system design approach that produces documents more suitable for LLM-based software development. Specifically, we generate only one document based on use cases. This document outlines the source files the desired system should have and the functions each file should implement.  

Concretely, in the designing phase, AISD receives the initial task and the use cases accepted by the user. It utilizes the prompts shown in Figure~\ref{fig:promDesign} to instruct LLMs in generating a list of source file names along with their corresponding descriptions. The system message in this prompt is designed by incorporating the ideas of role-playing~\cite{kong2023better} and few-shot learning~\cite{brown2020language} which we believe can enhance the quality of the outputs. Specifically, we have LLMs simulate the role of a software architect, tasking them with generating responses that are not only specific but also aligned with the given context. To facilitate the follow-up procedures, the outputs are expected to be in JSON format. To this end, we provide an output example in the prompt to demonstrate the expected outputs following the idea of few-shot learning. 

Figure~\ref{fig:pmpdesign} presents an example of the system design based on the revised use cases shown in Figure~\ref{fig:usecse2}. The system design is articulated through four Python source files, each accompanied by a concise explanation detailing its functions. We remark that this form of system design is simple yet LLMs-friendly, enabling LLMs to focus more on tractable sub-tasks by reducing irrelevant information. 

We remark that the system design involves a certain level of expertise, which poses challenges for ordinary users to review and modify it. Specifically, users often need to possess a deep understanding of programming languages and have some development experience to assess the reasonability of each module in the presented system design, as well as to ensure that all modules can form a complete system. Nevertheless, we enable users to determine if the generated system design is acceptable and opt to revise it, particularly when interacting with professional users. 

\begin{figure}
    \begin{ShortBox}
    "main.py": "This is the main file of the neural network classifier tool.",\\
    "classifier.py": "This file contains the implementation of the machine learning classification algorithm.",\\
    "gui.py": "This file provides the graphical user interface for users to enter iris characteristics and view classification results.",\\
    "utils.py": "This file contains utility functions used in the system."
    \end{ShortBox}
    \caption{Example of the system design in AISD}
    \label{fig:pmpdesign}
\end{figure}

\subsection{Coding and Automatic Testing}
\begin{figure*}[t]
    \begin{FullBox}
    \textbf{System Message:}\\    
    You are a Programmer. You have extensive computing and coding experience in many programming languages and platforms, such as Python.\\
    \textbf{User Message:}\\
    The user's task, use cases and original system designs are listed below:\\ 
    Task: "\{task\}".\\
    Use Cases: "\{use\_cases\}".\\
    System Design: "\{system\_design\}".\\
    You have to complete the task through an executable software with multiple files implemented via Python.\\ 
    To satisfy the new user's demands, you should write files and make sure that every detail of the architecture is, in the end, implemented as code. 
    The software should be equipped with graphical user interface (GUI) so that user can visually and graphically use it; so you must choose a GUI framework (e.g., in Python, you can implement GUI via tkinter, Pygame, Flexx, PyGUI, etc,).\\ 
    Think step by step and reason yourself to the right decisions to make sure we get it right. You will output the content of each file including complete code. Each file must strictly follow a markdown code block format, where the following tokens must be replaced such that "FILENAME" is the lowercase file name including the file extension, "LANGUAGE" in the programming language, "DOCSTRING" is a string literal specified in source code that is used to document a specific segment of code, and "CODE" is the original code: FILENAME LANGUAGE ~ \textquotesingle \textquotesingle \textquotesingle DOCSTRING\textquotesingle \textquotesingle \textquotesingle ~CODE You will start with the "main" file, then go to the ones that are imported by that file, and so on. Please note that the code should be fully functional. Ensure to implement all functions. No placeholders (such as `pass' in Python).
    \end{FullBox}
    \vspace{-0.1cm}
    \caption{Prompts used for code generation}
    \vspace{-0.1cm}
    \label{fig:pmpCode}
\end{figure*}

The code generation in AISD is an automatic and iterative process. With the system design generated previously, AISD prompts LLMs to produce all the required source files at once, tests the resultant system, and refines the source code according to the testing outcome. 

Prompts used to generate source code are shown in Figure~\ref{fig:pmpCode}. In the system prompt, we set the LLM to be a skilled programmer with experience in various programming languages. In the user prompt, we provide the initial task along with the outputs of each previous phase, i.e., use cases and system design. We then explain what LLMs should complete. Note that involving the task description and the corresponding use cases in the prompt is necessary, as these two types of information help LLMs retain the user's requirements, guiding them to code accordingly.

Instead of generating each code file separately (i.e., one code file in one chat session), the LLM is asked to complete all the coding tasks in one chat session. Typically, different source files collaborate to form the complete system, indicating that these files are interdependent. In practice, when generating code for interconnected modules in separate chat sessions, the system might encounter the challenge of ``robbing Peter to pay Paul'' as generating one file without considering the entire system can inadvertently result in a failure. This is because the generated code files may not coordinate with each other. Thus, in AISD, we instruct LLMs to generate all code files in one chat session, ensuring that the resultant files can work closely together, thereby reducing potential bugs in the system. Note that due to the hallucination of LLMs, there may exist some unimplemented functions (e.g., the function body just has a `pass' keyword in Python) and missing import packages in the resultant implementation. To remedy this, once the source code is generated, AISD employs LLMs to refine the source code, identifying and addressing these potential issues.

Limited by the capability of LLMs, the resultant implementation may still contain bugs even after automatic refinement with LLMs. Therefore, we introduce an automatic testing phase before the system is presented for human testing. During the automatic testing, AISD sequentially performs two different types of testing: unit testing, and system testing. For the unit testing, AISD generates a set of unit tests for each function, automatically runs them and records the results. Failure tests and the corresponding code are provided to LLMs for code refinement. Once all unit tests pass, AISD proceeds to the system testing, checking if the system can correctly start execution as a complete system. If errors occur during startup, AISD instructs LLMs to refine the code according to the error messages. Note that each testing is iterative, and the whole automation testing is implemented as an agent capable of automatically performing two kinds of testing and deciding when each testing stops. Additionally, we allow users to set a maximum number of iterations for each testing to prevent endless refinement. 

We take an example to showcase how AISD automatically refines the generated system through basic system testing. This example is observed when performing the development task ``Airplane War Game''. Specifically, when AISD executes ``python main.py'' for the system testing, an error occurs. The error message is shown in Figure~\ref{fig:sec33error} which suggests that a function is used wrongly. Subsequently, AISD employs the prompt shown in Figure~\ref{fig:sec33debugp} to instruct the LLM to fix the error accordingly. Note that the two placeholders ``{code}'' and ``{message}'' in that prompt are automatically replaced with the source code of the entire system implementation and the error message obtained at runtime.  

\begin{figure}
    \centering
    \begin{subfigure}{0.45\textwidth}
        \begin{lstlisting}[columns=fullflexible,numbers=none, framesep=5pt]
        Traceback (most recent call last):
        File "main.py", line 3, in <module>
              game.start_game()
        File "game.py", line 21, in start_game
          player.handle_input(player_airplane, bullets)
        TypeError: handle_input() missing 1 required positional argument: 'canvas'
        \end{lstlisting}
    \vspace{-0.1cm}
    \caption{Error message when running `python main.py'}
    \vspace{0.1cm}
    \label{fig:sec33error}
    \end{subfigure}
    \begin{subfigure}{0.45\textwidth}
    \begin{FullBox}
    \textbf{System Message:}\\
    Please review the source code. Identify and fix the issues listed below. Think step by step. First, analyze the reason why errors are increasing. Second, write the entire code files, ensuring that the format matches that of the Source Code. 
    Each file must strictly follow a markdown code block format, where the following tokens must be replaced such that "FILENAME" is the lowercase file name including the file extension, "LANGUAGE" in the programming language, "DOCSTRING" is a string literal specified in source code that is used to document a specific segment of code, and "CODE" is the original code: FILENAME LANGUAGE ~ \textquotesingle \textquotesingle \textquotesingle DOCSTRING\textquotesingle \textquotesingle \textquotesingle ~CODE You will start with the "main" file, then go to the ones that are imported by that file, and so on. Please note that the code should be fully functional. Ensure to implement all functions. No placeholders (such as "pass" in Python).\\
    \textbf{User Message:}\\
    Source Code: \{code\}\\
    Problem: \{message\}  
    \end{FullBox}
    \vspace{-0.1cm}
    \caption{Prompts used for fixing runtime bugs}
    \label{fig:sec33debugp}
    \end{subfigure}
    \vspace{-0.1cm}
    \caption{Example of automatic testing}
    \vspace{-0.5cm}
\end{figure}
\subsection{System Validation}
\begin{figure}[h]
    \begin{subfigure}{0.4\textwidth}
        \centering
        \includegraphics[width=0.8\linewidth]{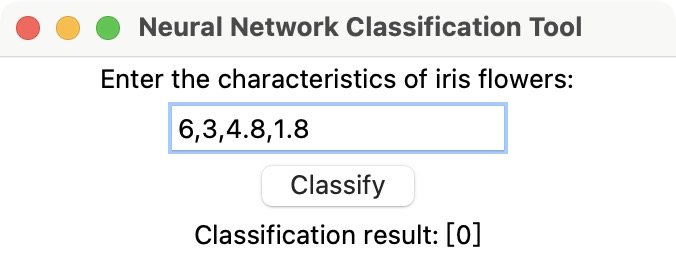}
        \caption{The resultant system before manual testing}
        \label{fig:iris1}
        \vspace{0.1cm}
    \end{subfigure}
    \begin{subfigure}{0.4\textwidth}
        \centering
        \includegraphics[width=0.8\linewidth]{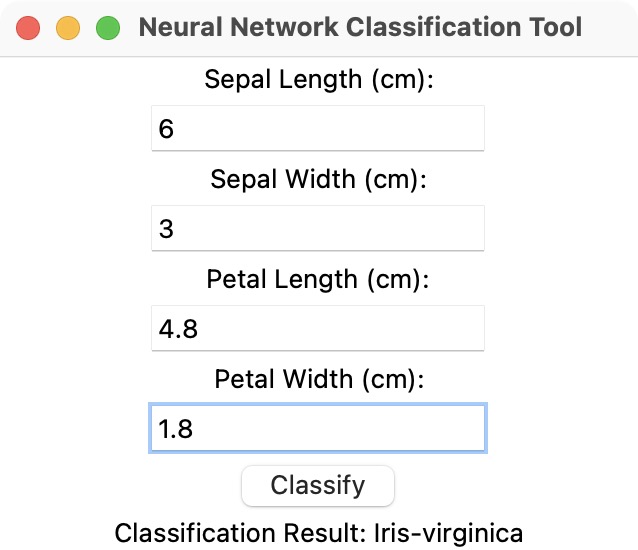}
        \caption{The resultant system after manual testing}
        \label{fig:iris2}
    \end{subfigure}
    \caption{Comparison of the resultant system before and after manual testing}
    \label{fig:iris_comparison}
\end{figure}

Passing the unit testing and being executable do not necessarily mean that the generated system is acceptable from the perspective of stakeholders. The main reason is that the automatic testing, as depicted earlier, is unable to verify if the functional requirements are satisfied, especially when 
the initial set of use cases may be still incomplete or wrong according to the actual user expectation, or the system is built with a graphic user interface (for which there will be typically a lot of details that may not be what the user wants). For example, the use case ``User can view the classification results on a board'' in Figure~\ref{fig:usecse2} directly requires the user's feedback. In this case, it is challenging to assess whether this requirement is satisfied without human engagement.

To ensure that the generated system aligns with the user's requirements, it is essential to involve the user in validating the system. Specifically, when the generated system passes the automatic testing, the user is required to manually test the system according to the use cases generated previously. If the requirement depicted by a use case is not satisfied, the user is then required to revise the description of the use case or instruct AISD to revise the source code. Concretely, if the user fails to validate a use case but no errors occur when validating it, the user then revises the description of the use case to make it more detailed and instructs AISD to start from the system designing phase again. In case of errors reported, the user then provides the error message and instructs AISD to fix the bug accordingly. Lastly, the user can also choose to introduce new use cases (i.e., new requirements), which is often the case in practice. The user can repeat this process until all the use cases pass the testing. 

We present an example showcasing how the user collaborates with AISD in this phase. Figure~\ref{fig:iris1} shows the resultant system based on the use cases presented in Figure~\ref{fig:usecse2}. However, this system does not meet the user's expectations. Firstly, there is only one input box with no prompts on how to enter the iris's characteristics. Secondly, the classification result is presented as a number, which is hard for users to understand. The second issue is interesting which exposes that humans and LLMs may have different understandings about the same expression, e.g., ``Use can view the classification result''. While users indeed can ``view'' the classification result, they generally expect the tool to directly present what species the input iris is, instead of a cryptic number (though it may correspond to a specific iris species). Therefore, we revise the use cases to add some details. The revised version is shown in Figure~\ref{fig:usecse3}. Subsequently, we instruct AISD to proceed with the updated use cases. After a series of steps as described previously, the resultant system aligns perfectly with our requirements, as shown in Figure~\ref{fig:iris2}. 

\section{Experiments}
\label{sec:4}
We have implemented AISD as a self-contained toolkit based on LLMs, comprising approximately 2058 lines of source code written in Python 3.7. Our implementation as well as the benchmark used in this work are available at~[\citenum{ourcode}]. In the following, we conduct multiple experiments to address the following research questions:
\begin{itemize}
  \item RQ1: How effective is AISD in developing software?
  \item RQ2: Does the human engagement matter? 
  \item RQ3: How much user involvement is needed for AISD to work effectively?
\end{itemize}
RQ1 aims to evaluate the effectiveness of AISD in completing development tasks. RQ2 further conducts an ablation study to verify if human engagement does contribute to the effectiveness of AISD. Finally, RQ3 analyzes the number of user interactions required by AISD when performing a software development task. 

\subsection{Experimental setup}
\noindent \textbf{\emph{Assessment benchmark.}} We have manually constructed a benchmark named CAASD (\textbf{C}apability \textbf{A}ssessment of \textbf{A}utomatic \textbf{S}oftware \textbf{D}evelopment) to assess the capabilities of various AI-aided software development systems, which we consider to be another contribution of this work.  

Note that existing benchmarks are not suitable for evaluating the capability of the AI-aided software development system. This is because benchmarks like HumanEval~\cite{humanEval}, MBPP~\cite{mbpp} and CAMEL~\cite{li2023camel} either consist of simple function-level coding tasks or lack relatively objective criteria to assess the level of task completion. The former essentially reflects the capability of the LLM itself (since LLMs are effective for solving such tasks), while the latter hinders an objective assessment of task completion. To address the problem of lacking a benchmark for evaluating LLM-based automatic software development systems, we propose the CAASD benchmark which contains 72 software development tasks collected from multiple sources and from multiple domains such as small games, personal websites, and various other applications. On average, the implementation of each task in CAASD requires approximately 240 lines of code, based on the analysis of open-source implementations of similar tasks~\cite{geek, openBMB, gamesdata}. Each test case in CASSD consists of four fields, i.e., a task ID, a task name, a task prompt and a list of reference use cases. Particularly, the task prompt describes the task to complete (i.e., a high-level often vague system requirement) and is provided as input to various AI-aided software development systems. The reference use cases of each task depict the essential functional requirements of a system. We remark that these reference use cases serve as a means to objectively assess the task completion, details of which will be elaborated in Section~\ref{sec:4.2}. \\

\noindent \textbf{\emph{Baselines.}} We compare AISD with two state-of-the-art baselines, namely ChatDev~\cite{chatDev} and MetaGPT~\cite{metagpt}. ChatDev divides the development process into four phases (i.e., designing, coding, testing, and documenting), and assigns two virtual roles to solve the corresponding subtask at each phase by multi-turn discussions. The two roles stop discussing only when they reach a consensus that the subtask has been successfully accomplished. MetaGPT also decomposes a general development task into several subtasks, but differs from ChatDev in two aspects. First, when solving each subtask, MetaGPT adopts only one virtual role (e.g., a product manager) to address the subtask. Second, MetaGPT incorporates the idea of human Standardized Operating Procedures (SOPs) and mandates each role to produce standardized outputs, facilitating knowledge sharing across different modules.  

We conducted all experiments based on ChatGPT~\footnote{https://openai.com/blog/introducing-chatgpt-and-whisper-apis} with version ``gpt3.5-turbo-16k''. For ChatDev, we utilize the settings outlined in~\cite{chatDev}, whereas for the MetaGPT, we adopt the default settings of the open-sourced implementation\footnote{https://github.com/geekan/MetaGPT(v0.3.0)} as the paper~\cite{metagpt} contains insufficient details of the experimental settings. Following the settings in ChatDev, we allowed up to 5 turns for both automation testing and manual testing in AISD, and selected the best as the final outcome among the systems generated from all turns.

\subsection{Research Questions}\label{sec:4.2}

\noindent \emph{\textbf{RQ1: How effective is AISD in developing software?}} To answer this question, we applied AISD and the two baselines to solve the tasks of the CAASD benchmark systematically, and then reported the pass rate of the tasks and costs for each approach. The ``costs'' here refers to the overall count of tokens consumed by LLMs to accomplish a task, encompassing both input tokens and output tokens. The pass rate for each task is calculated using the following formula:
\[
  \textit{Pass rate} = \frac{\textit{\#Passed Use Cases}}{\textit{\#Total Use Cases}}
\]
where \textit{\#Total Use Cases} is the total number of the reference use cases of a task in CAASD, and \textit{\#Passed Use Cases} denotes the number of reference use cases which pass the manual testing. Specifically, we manually inspect reference use cases of the corresponding task one by one by running the generated system. The value of \textit{\#Passed Use Cases} is then obtained by counting the number of successfully passed use cases.

\begin{figure}[t]
  \centering
  \includegraphics[width=0.48\textwidth]{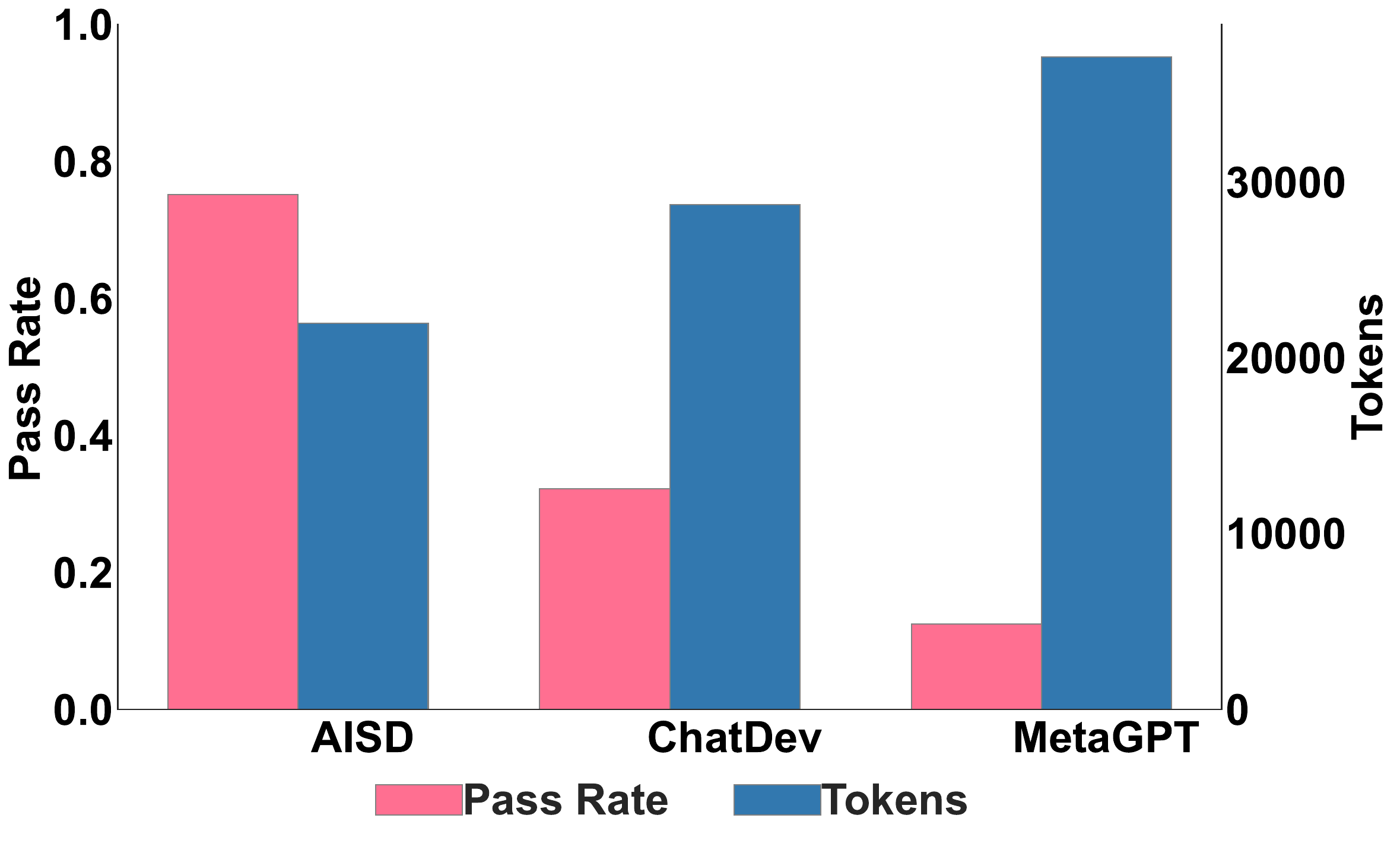}
  \caption{The performance of three AI-aided software development systems on CAASD benchmark }
  \label{fig:rq1}
\end{figure}

The results are summarized in Figure~\ref{fig:rq1}. We observe that our approach, i.e., AISD achieved an impressive \mkchange{75.2}\% pass rate while maintaining an average token consumption of just \mkchange{21993} tokens per task. In contrast, ChatDev has a pass rate of \mkchange{32.2}\% and consumes an average of \mkchange{28734} tokens per task, and MetaGPT performs even worse with a pass rate of only \mkchange{12.5}\% and an average token cost of \mkchange{37136} per task. It is worth noting that both MetaGPT and ChatDev have a cost that is over \mkchange{1.3} times higher than that of AISD. The significant improvement of AISD over ChatDev may be attributed to the human engagement in its design (which we will systematically analyze in RQ2). MetaGPT is dramatically less effective compared to AISD and ChatDev, although it is not surprising. The reason is that MetaGPT feeds too much complex information to LLMs, leading to difficulties for LLMs to understand the tasks. For example, before the coding phase, MetaGPT generates a standard PRD (Product Requirements Document), including ``Product Goals'', ``User Stories'', ``Competitive Analysis'', ``Competitive Quadrant Chart'' (in the form of mermaid\footnote{https://github.com/mermaid-js/mermaid} code), ``Requirement Analysis'', ``Requirement Pool''. In the coding phase, the PRD with this rich information is directly used as a part of prompt to instruct LLMs, which may result in cognitive overload for LLMs. The results from MetaGPT further reinforce the point that we highlighted in Section~\ref{sec:32} (i.e., intricate and overwhelming information potentially distracts LLMs), ultimately undermining their ability to comprehend and complete core tasks. Additionally, among the three frameworks, AISD consumed the fewest tokens, while MetaGPT consumed the most, indicating that AISD is not only effective but also efficient. 

\begin{framed}
  \noindent \emph{Answer to RQ1: AISD significantly improves the use cases pass rate with lower costs compared to existing baselines.}
\end{framed}
\noindent \emph{\textbf{RQ2: Does the human engagement matter?}}

\begin{figure}[t]
  \centering
  \includegraphics[width=0.45\textwidth]{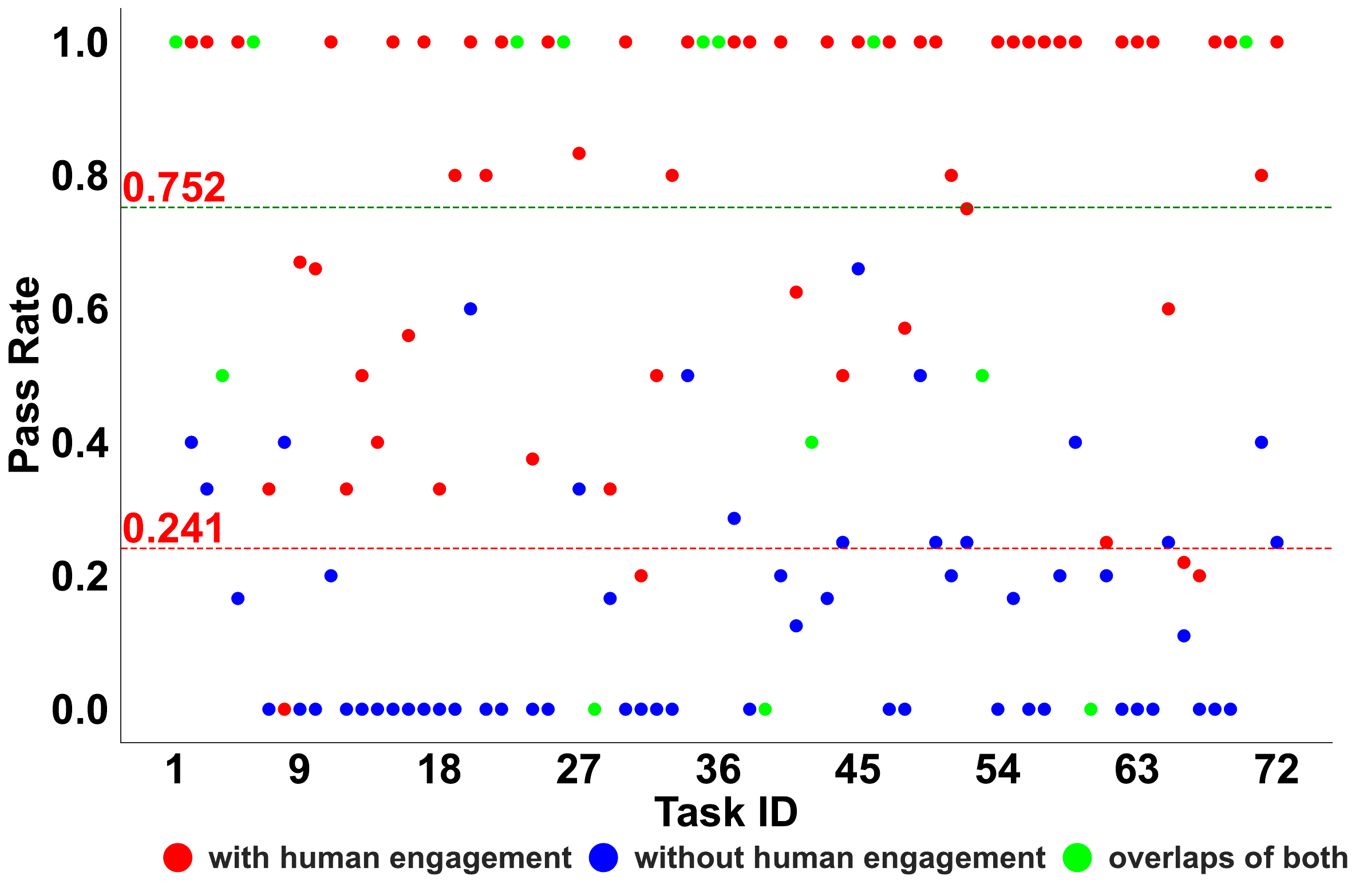}
  \caption{Pass rate of AISD with and without human engagement over CAASD benchmark.}
  \label{fig:rq21}
\end{figure}
To address this question, we applied AISD to the CAASD benchmark again but omitted the human engagement in all phases. That is, we skipped the manual revision in the use cases generation and the entire manual testing phase. Note that we conducted 5 trials for each task, and selected the best pass rate among the 5 trials as the pass rate of the task. We compared the results obtained without human engagement to those with human engagement. 

Figure~\ref{fig:rq21} illustrates the pass rate of AISD with and without human engagement over CAASD benchmark. The x-axis represents the ID of each task, ranging from 1 to 72. For each task, the pass rate with human engagement is denoted by a red point, and the pass rate without human engagement is denoted by a blue point. The green point indicates that there is an overlap of a blue point and a red point. The green horizontal line and the red horizontal line represent the average pass rate of each group, respectively. Observing the data, the average pass rate of AISD with human engagement (i.e., about \mkchange{75.2}\%) is approximately \mkchange{51.1} percentage higher than that of the AISD without human engagement (i.e., about \mkchange{24.1}\%). It is worth noting that a concentration of \mkchange{47.2}\% orange points is evident at the bottom (i.e., 0\% pass rate), contrasting with only \mkchange{5.6}\% of blue points at the same level. These findings suggest that human engagement not only enhances the overall use case pass rate of the generated system implementation, but also underscores the significance of human involvement in handling tasks that prove challenging when relying solely on the power of LLMs. In the next research question, we will conduct a more in-depth analysis of the influence of human involvement on the use case pass rate.

\begin{framed}
  \noindent \emph{Answer to RQ2: Engaging users during requirement analysis and system testing effectively bridges the gap between the generated system and users' expectation, especially in handling challenging tasks.}
\end{framed}
\begin{table}[t]
  \caption{The instances of revisions achieving the final pass rate using AISD.}
  \label{tab:rq3t1}
  \begin{tabular}{ccccc}
  \toprule
  $h_1$ & $h_2$ & \#Task  & Avg. Pass Rate (\%)  & Total Revisions  \\
  \hline
     0 &    0 &  1  &100\%   & 0 \\
     0 &    1 &  1  &100\%   & 1  \\
     0 &    2 &  0  &-       & - \\
     0 &    3 &  0  &-       & - \\
     0 &    4 &  0  &-       & - \\
     0 &    5 &  0  &-       & - \\
     1 &    0 &  14 &68.57\%       & \mkchange{1} \\
     1 &    1 &  8  &81.44\%  & 2 \\
     1 &    2 &  12 &\mkchange{85.42}\%  & 3 \\
     1 &    3 &  11 &86.68\%  & 4 \\
     1 &    4 &  9 &86.89\%  & 5 \\
     1 &    5 &  16 &52.59\%  & 6 \\
     \hline
    \multicolumn{4}{c}{Average times of manual revision} & 3.5\\
     \bottomrule
  \end{tabular}
  \end{table}
  \begin{figure}[t]
    \centering
    \includegraphics[width=0.45\textwidth]{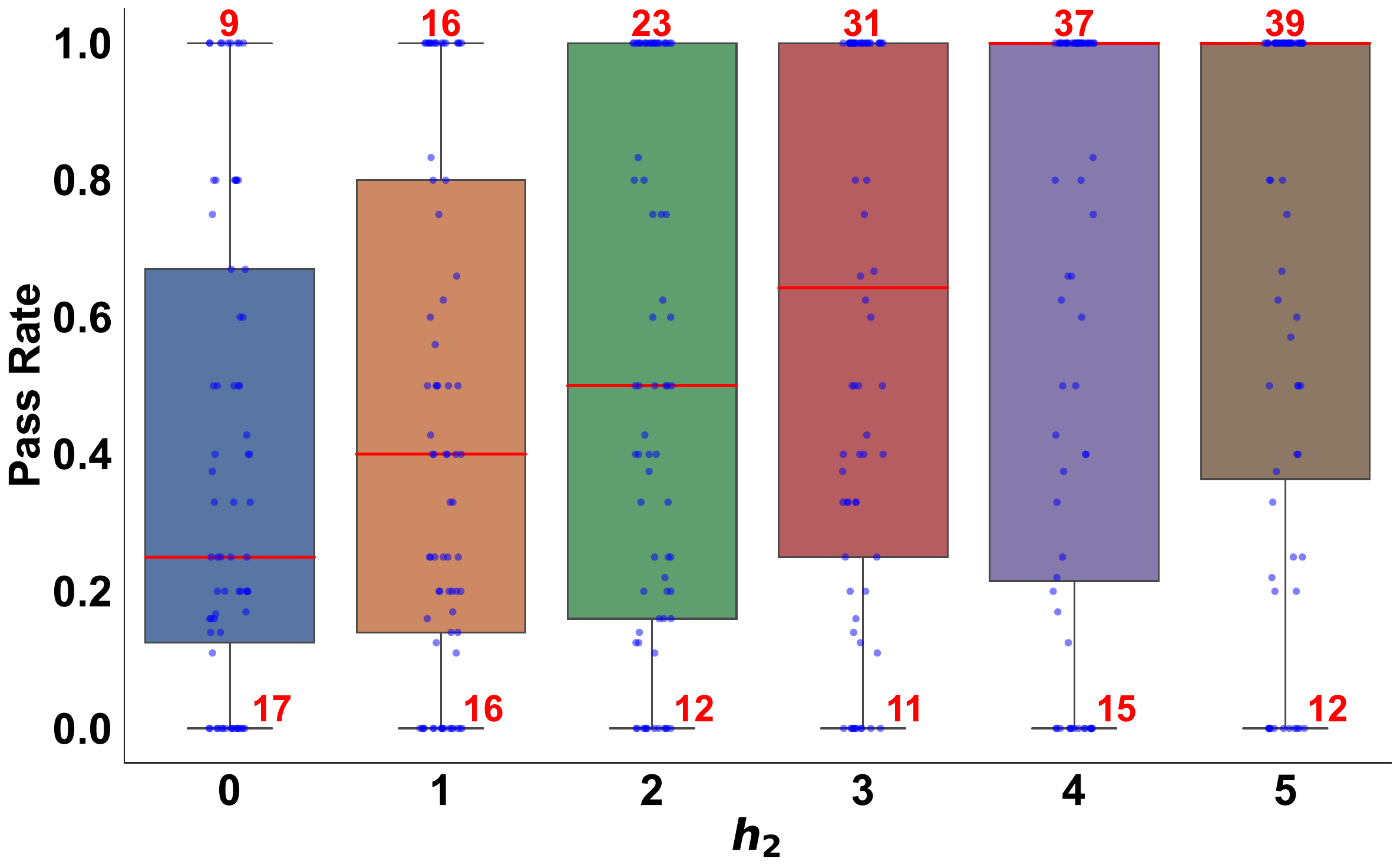}
    \vspace{-0.2cm}
    \caption{The pass rate with different numbers of human interactions ($h_2$)}
    \vspace{-0.4cm}
    \label{fig:rq3f1} 
  \end{figure}
\noindent \emph{\textbf{RQ3: How many times of user involvement are needed in AISD?}} In AISD, users mainly engage in two distinct phases. The first involves use case generation, and the second entails manual testing. In the former phase, users are required to review the generated use cases and revise them if necessary until they are deemed to be acceptable. In the latter phase, users participate in an iterative process wherein they assess the use cases by executing the resultant system, and they may subsequently revise the use cases or prompt LLMs to address bugs. Consequently, users are involved in the entire development process at least twice, even if they choose not to revise the use cases or perform bug fixing based on error messages. To show the required manual efforts using AISD more specifically, we counted the times of manual revision in both use cases generation and the manual testing phases for each development task. In this context, ``manual revision'' refers to the manual adjustment of use cases or the prompt for fixing bugs.

Table~\ref{tab:rq3t1} shows the manual revisions involved in completing development tasks with AISD. In this table, $h_1$ denotes the times that a human revises use cases during the use case generation phase, $h_2$ denotes the times that a human revises use cases or prompts LLMs to fix bugs during the manual testing phase, `\#Task' is the number of tasks each of which undergoes $h_1$ revisions during the use case generation phase and $h_2$ revisions during the manual testing phase, `Avg. Pass Rate' indicates the average pass rate achieved by tasks of each group, and `Total Revisions' denotes the total times of manual revisions of each task, which is the sum of $h_1$ and $h_2$. Note that $h_1$ is either 0, indicating no modifications, or 1, indicating there are manual revisions during the use case generation phase. Note that all modifications are treated as one revision since there is no iteration during this phase. The value of $h_2$ ranges from 0 to 5, as we allow a maximum of 5 iterations during the manual testing phase in this work. 

We can observe that AISD requires approximately four revisions on average. For over half of the tasks (47 out of 72), AISD achieves the best pass rate within four rounds of manual revision. Note that only one task out of the total 72 tasks attains a 100\% pass rate without any manual revisions, and 70 tasks (where $h_1$ equals 1) demonstrate the need for user involvement in revising the generated use cases (i.e., the generated use cases look implausible and thus need manual modifications). These findings underscore a substantial disparity between use cases generated by LLMs and user expectations, emphasizing the essential role of human engagement in successfully completing development tasks. 

We further analyze how the pass rate varies with the increasing number of human revisions. Figure~\ref{fig:rq3f1} displays changes in key statistical measures of pass rate as $h_2$ increments through a box-and-whisker plot with observations. The numbers in red are the counts of minimum and maximum points. Note that the pass rate in this figure is obtained based on the resultant system after $h_2$\textit{-th} revision. In general, with the rise in $h_2$, the median pass rate and the percentiles (i.e., 25th and 75th) gradually rise, and tasks attaining a 100\% pass rate (depicted by blue points at the top) become more concentrated. The median pass rate even reaches 100\% when $h_2>3$. These findings, to some extent, rule out the possibility that the pass rate improvement is solely a result of the uncertainty of LLMs. We also observe that when increasing $h_2$, there are still some tasks for which the pass rate remains 0\%. This is not surprising. First, the quality of the revision depends on both the user's proficiency and the task's complexity. When handling particularly complex tasks, it becomes challenging for users to discern the required functions of the system, hindering their ability to revise use cases effectively. Additionally, some use cases surpass the capabilities of LLMs. For instance, in the development task ``Voice Assistant'', users expected that the desired system can understand what they said, and respond correctly. The system delivered by AISD however can only record the voice of users, even after rounds of prompting. 
\begin{framed}
  \noindent \emph{Answer to RQ3: In general, increasing user interactions leads to a higher pass rate of use cases. However, for some complex tasks, additional interactions may yield limited results due to LLMs' constraints. Our practical guideline is thus to continue interacting with AISD until users struggle to improve the generated use cases, and no runtime error occurs.}
\end{framed}
\vspace{-0.5cm}
\section{Related Work} \label{sec:5}
This work is closely related to existing approaches on automatic code generation. Automatic code generation is a hot topic in natural language processing community. Before the emergence of large neural network models, the works on this topic can be categorized into two groups. One primarily focuses on traditional techniques from both rule-based and statistical natural language processing, while the other is related to neural networks. 

In the early stages, researchers primarily employ heuristic rules or expert systems to synthesize program. Jha et al.~\cite{jha} propose an approach to automatically synthesize loop-free programs based on a combination of oracle-guided learning from examples and constraint-based synthesis from components using satisfiability modulo theories (SMT) solvers. Allamanis and Sutton~\cite{allamanis2014mining} propose to extract code idioms from a corpus of idiomatic software projects by nonparametric Bayesian probabilistic tree substitution grammars. Raychev et al.~\cite{raychev2014code} synthesize code completions by identifying the highest-ranked sentences with a statistical language model. However, these approaches have limitations in scalability and are confined to simple scenarios. For example, they struggle to generate a complete program from scratch according to a natural language description. 

Subsequently, with the rise of deep learning, neural networks including convolutional neural networks (CNNs)~\cite{li2021survey} and various recurrent neural networks (RNNs)~\cite{lipton2015critical} are widely adopted to approach the natural-language-to-code task. Ling et al.~\cite{ling2016} treat the code generation task as a sequence-to-sequence problem, and then propose an LSTM-based neural network architecture to generate code from natural language. Sun et al.~\cite{sun2019grammar} design a grammar-based structural CNN to generate a program by predicting the grammar rules of the programming language. Although these works achieve significant improvement in flexibility and scalability, they still exhibit poor generalization ability due to limited language understanding. 

After the introduction of the Transformer~\cite{vaswani2017attention}, numerous large language models (LLMs) have been proposed and have demonstrated impressive results in automatic code generation. For example, Codex~\cite{humanEval} can solve 72.31\% of challenging Python programming problems created by humans. We remark that while LLMs exhibit remarkable proficiency in programming, instructing them directly to complete complex software development tasks is challenging. In response, researchers have put forth a series of LLM-based multi-agent frameworks to automatically generate software with an initial idea.

Li et al.~\cite{li2023camel} propose \emph{role-playing}, a communicative agent framework, to complete a task (not limited to software development tasks) by having three agents communicate with each other. Specifically, when the human user specifies a task to implement, a task specifier agent specifies a role to play for the assistant agent and user agent respectively, and then the two roles engage in conversation to complete the task. ChatDev~\cite{chatDev} divides the development process into four stages: designing, coding, testing, and documenting. Each stage involves a group of agents to solve a distinct subtask, such as deciding the software modality and programming language at the designing phase. Note that the subtask of each stage is further decomposed into atomic subtasks, each of which is addressed by two agents engaging in conversation. Inspired by the human workflows, MetaGPT incorporates Standardized Operating Procedures (SOPs) into their framework to coordinate agents. MetaGPT also mirrors the waterfall model and breaks down the development process into several phases. The main difference is that each agent in MetaGPT solves a subtask independently and produces standardized action outputs for knowledge sharing. We remark that these existing frameworks exclude humans from the development process, and relying solely on the abilities of LLMs may undermine their effectiveness when facing complex tasks. 
\section{Conclusion} 
\label{sec:6}
In this work, we present an AI-powered software development framework. Different from existing approaches, our framework is designed to keep users lightly engaged when solving a complex task, emphasizing the importance of human-engaged requirement analysis and system validation. To objectively assess the capabilities of completing software development tasks, we have built a novel benchmark, each task of which is equipped with a list of use cases for reference during assessment. The evaluation results demonstrate that our framework significantly improves the task pass rate while consuming fewer tokens.
\bibliographystyle{ACM-Reference-Format}
\bibliography{ref}
\end{sloppypar}
\end{document}